\documentclass[11pt]{article}

\usepackage{amsmath}
\usepackage{amscd}
\usepackage{amssymb}
\usepackage{amsfonts}
\usepackage{eufrak}
\usepackage[mathscr]{euscript}
\usepackage{pstricks}
\usepackage{pst-node}
\usepackage[dvips]{pstcol}\textheight=8.5truein
\def \gt{\gamma_t}
\def \ggt{\gamma_t \otimes \gamma_t}
\def \git{\gamma_t \otimes {\bf I}_n}
\def \gto{\gamma_t^{(1)}}
\def \gtt{\gamma_t^{(2)}}
\def \ggtot{\gamma_t^{(1)} \otimes \gamma_t^{(2)}}
\def \Tr{{\rm Tr}}
\def \mn{M_n({\bf C})}
\def \mmn{M_n({\bf C}) \otimes M_n({\bf C})}

\def \mt{M_2({\bf C})}
\def \mmt{M_2({\bf C}) \otimes M_2({\bf C})}

\def \ggtote{\gamma_t \otimes \gamma_t^{\varepsilon}}

\begin{document}

\title{\bf Complete positivity and dissipative factorized dynamics}
\author{Fabio Benatti$^{a,b}$, Roberto Floreanini$^b$, Marco
Piani$^{a,b}$, Raffaele Romano$^{a,b}$\footnote{E-mail address: rromano@ts.infn.it}
	\\ $^a$Dipartimento di Fisica Teorica \\ Universit\`a di Trieste
\\ Italy 
        \\ $^b$ Istituto Nazionale di Fisica Nucleare \\ Sezione di
Trieste \\ Italy}
\date{Lisboa, Portugal, July 2003}
\maketitle

\begin{abstract}

\noindent
After reviewing the main properties of time-evolutions of open quantum
systems, some considerations about the positivity of
factorized Markovian dynamics for bipartite
systems are made. In particular, it is shown that the
positivity of the whole time-evolution in general does not ask for
the complete positivity of the single system time-evolutions, if
they are allowed to differ.
However, they must be completely positive if one is a small
perturbation of the other, which is the typical situation for open
systems in a heat bath.

\end{abstract}


\section{Introduction}

The interaction of a quantum system with its surroundings is a source
of irreversibility in its time-evolution: the usual unitary
dynamics, given by the Schr{\" o}dinger equation, has to be replaced
by a more general dynamics irreversible and producing
decoherence, in general not-Markovian~\cite{ali,bre}. 
Under some rather broad
assumptions the memory terms can be neglected, producing a
time-evolution with semigroup structure. The form 
of such a time-evolution is uniquely fixed
(Lindblad~\cite{lin}; Gorini, Kossakowski, Sudarshan~\cite{gor}) 
and it characterizes a wide
variety of physical phenomena ranging from quantum optics~\cite{qop1,qop2}, 
magnetic resonance~\cite{nmr}, statistical mechanics, 
foundational aspects of quantum mechanics~\cite{ghi} 
to elementary particle physics (see~\cite{raf} and references therein).

The consistency of the physical interpretation of the formalism
strongly relies on the property of complete positivity (in the
following CP) of the open system time-evolution~\cite{bre}. 
It turns out that this property, stronger
than simple positivity, is indeed necessary when dealing with
composite systems, that is systems built up from several possibly
not-interacting subsystems. The necessity of CP shows up 
when there is an initial entanglement between them~\cite{ben1,ben2}. 
In other words, the relevance of CP depends
on statistical, not dynamical, properties of the considered composite
system; for this reason we limit our attention to factorized dynamics,
characterizing two not-interacting subsystems. 
In the Markovian case it has been shown that the
positivity of the factorized map $\ggt$ is a necessary and sufficient
condition for the map $\gt$~\cite{ben3} being completely positive. 
In this contribution we
explicitly show that this result does not generalize to
two subsystems evolving under different
Markovian time-evolutions $\gto$
and $\gtt$: positivity of $\ggtot$ does not imply in general 
that both $\gto$ and
$\gtt$ are completely positive. However, if $\gtt$ is a small
perturbation of $\gto$, CP of both $\gto$ and $\gtt$ is required.

In Section~\ref{s1} a brief introduction to open quantum systems and
their time-evolution is given, pointing out the assumptions leading
to a Markovian approximation of the dynamics. We will be concerned
only with this kind of time-evolutions. In
Section~\ref{s2} the properties of positivity and complete positivity
of the dynamics are reviewed, stressing out the relevance of CP and
its relation to quantum entanglement. Finally, in Section~\ref{s3} we
deal with factorized dynamics and discuss the main results of the
present work.


\section{Open quantum systems and their time-evolution}
\label{s1}

In the following S will denote the system we are interested in; it can
be either a well defined physical system or an abstract $n$
($\leqslant + \infty$)
level system. The outer system, called environment, will
be denoted by E.

{\bf Definition 1.} S is called {\it closed} if there is no
interaction between S and E; otherwise {\it open}.

Physical states of S will be represented by statistical operators (or
density matrices) $\rho_S$, that is positive operators with unit trace
acting on the Hilbert space of S, ${\cal H}_S$:

\begin{equation}
\label{rho}
\rho_S = \rho_S^{\dagger}, \quad
\rho_S \geqslant 0, \quad
\Tr (\rho_S) = 1.
\end{equation}

The statistical
operators can be either pure (one-dimensional projectors) or mixed 
(convex linear
superpositions of pure states); their spectral decomposition reads

\begin{equation}
\label{rho2}
\rho_S = \sum_i \lambda_i \vert \psi_i \rangle \langle \psi_i \vert
\end{equation}

\noindent
where $\{\vert \psi_i \rangle\}$ is a basis for ${\cal H}_S$ and the
eigenvalues $\lambda_i$ represent the weights of the statistical
superposition; in fact, $\lambda_i \geqslant 0$ and $\sum_i \lambda_i
= 1$, as a consequence of~(\ref{rho}). Then~(\ref{rho}) are fundamental
in order to have a consistent statistical interpretation of the
density matrices formalism which assignes to the $\lambda_i$ the role
of probabilities. 
The observables subject to measurement are
represented by Hermitian and bounded linear operators $A =
A^{\dagger}$ and the mean value of $A$ over the state $\rho_S$ is
given, in this formalism, by the trace operation:

\begin{equation}
\label{mean}
\langle A \rangle_{\rho_S} = \Tr (A \rho_S).
\end{equation}

{\bf Definition 2.} By {\it time evolution} we mean a continuous  
one-parameter
family of linear operators, $\{ \gt \}$, mapping states into states:
$\rho_S (t) = \gt [\rho_S (0)]$, where $\rho_S (0)$, $\rho_S (t)$ are the
statistical operators representing the system at the initial and
at a later time $t$, respectively.

For example, the time-evolution of a closed system S reads

\begin{equation}
\label{closed}
\rho_S(t) = \gt [\rho_S(0)] = U_t \rho_S(0) U_t^{\dagger}, \quad U_t =
e^{-i H_S t}
\end{equation}

\noindent
where $H_S$ is the Hamiltonian of S. In differential form we get the
Liouville-von Neumann equation

\begin{equation}
\label{closed2}
\dot{\rho}_S(t) = L_H [\rho_S(t)] = -i[H_S, \rho_S(t)]
\end{equation}

\noindent
generated by the Liouville superoperator $L_H [\,\cdot \,] = -i
[H_S,\,\cdot \,]$.

If we assume the total system T = S + E to be closed, the time-evolution for
the open system S is obtained tracing over the environment degrees of
freedom: 

\begin{equation}
\label{open}
\rho_S(t) = \gt [\rho_S(t)] = \Tr_E [U_t \rho_T(0) U_t^{\dagger}],
\quad U_t = e^{-i H_T t},
\end{equation}

\noindent
where $H_T = H_S + H_E + H_I$ is the Hamiltonian of the total system
T, $H_E$ the Hamiltonian of E and $H_I$ the interaction term. The
differential form of~(\ref{open}), called generalized master equation
(GME), is 

\begin{equation}
\label{open2}
\dot{\rho}_S(t) = -i [H_S^{\rm eff},\rho_S(t)] + \int_0^t {\cal K}(t - u)
\rho_S(u) du + {\cal I}(t).
\end{equation}

Some comments concerning~(\ref{open2}): $H_S^{\rm eff}= H_S + \Tr_E(H_I
\rho_E)$ is the redefined Hamiltonian of S, containing a Lamb-shift
term ($\rho_E$ is a reference state of E). The convolution term
in~(\ref{open2}) depends on the kernel ${\cal K}$ and contains the
whole history of $\rho_S$ from initial time to $t$ (memory term).  
The inhomogeneity ${\cal I}$ is
related to initial correlations between S and E
(explicit expressions of ${\cal K}$ and ${\cal I}$ can be found 
in~\cite{haa}). If we switch off the
interaction, $H_I = 0$, eq.~(\ref{open2}) reduces to the Liouville-
von Neumann equation since ${\cal K} = 0$, ${\cal I} = 0$ and also the
Lamb-shift term drops out. 

It is difficult to deal with the dynamics generated by~(\ref{open2})
because of the convolution term. However, in many physical
situations it is possible to approximate it with a Markovian dynamics,
that is a memoryless dynamics characterized by a linear generator $L$
that generalizes the Liouville superoperator:

\begin{equation}
\label{markov}
\dot{\rho}_S (t) = L [\rho_S (t)] = L_H^{\rm eff} [\rho_S(t)] + L_D
[\rho_S(t)].
\end{equation}

Besides the redefined Hamiltonian part $L_H^{\rm eff} [\,\cdot\,] = -i
[H_S^{\rm eff},\,\cdot\,]$, the additional contribution
$L_D$ has appeared. Its form is uniquely fixed by a theorem that in
the finite-dimensional Hilbert space case is as
follows~\cite{gor}:

{\bf Theorem 1.} For a finite dimensional Hilbert space (dim${\cal
H}_S = n$), $L_D$ in~(\ref{markov}) is given by

\begin{equation}
\label{kossak}
L_D [\rho_S] = \sum_{i,j = 1}^{n^2 - 1} c_{ij} \Bigl[F_i \rho_S
F_j^{\dagger} - \frac{1}{2}\Bigl\{ F_j^{\dagger} F_i, \rho_S 
\Bigr\} \Bigr]
\end{equation}

\noindent
where $\{F_i, i = 1, n^2 - 1\}$ is an orthonormal basis for the space
of traceless $n \times n$ matrices: $\Tr (F_i^{\dagger} F_j) =
\delta_{ij}$, $\Tr (F_i) = 0$ and $C = [c_{ij}]$ is a self-adjoint
matrix. 

The assumptions underlying
the Markovian approximation~(\ref{markov}) are: $a)$ weak interaction
between S and E; $b)$ absence of initial correlation: $\rho_T(0) = \rho_S(0)
\otimes \rho_E$. The generated time-evolution, $\gt [\,\cdot\,] =
e^{Lt}[\,\cdot\,]$, is then irreversible (since it is defined only for $t
\geqslant 0$), Hermiticity and trace-preserving, decohering (i.e. it
causes the transition from pure to mixed states). Does it, as it
must, preserve the positivity of states: $\rho_S(t) = \gt[\rho_S(0)]
\geqslant 0 \,\, \forall t \geqslant 0$? We shall shortly return to
this point in the next section. The one-parameter (time) family of
maps $\{ \gt, t \geqslant 0 \}$ is a semigroup of transformations
since it satisfies the forward in time composition law: $\gamma_t
\circ \gamma_s = \gamma_{t + s}\,\, \forall t,s \geqslant 0$.


\section{Positivity and complete positivity}
\label{s2}

For the sake of the statistical interpretation, the time-evolution $\gt$
we have described must be positivity-preserving.
Denoting by $\mn$ the algebra of $n \times n$ complex matrices,
for the map $\gt : \mn \rightarrow \mn$ we have the following
definitions:

{\bf Definition 3.} $\gt$ is said to be {\it positive} (or {\it
positivity-preserving}) if and only if it preserves the positivity of
the states it acts on:

\begin{equation}
\label{posit}
\gt [\rho_S] \geqslant 0 \quad \forall \rho_S \geqslant 0.
\end{equation}

A stronger property is as follows:

{\bf Definition 4.} $\gt$ is said to be {\it completely positive} (CP)
if and only if its lifting $\gamma_t \otimes {\bf I}_d : \mn \otimes
M_d({\bf C}) \rightarrow \mn \otimes M_d({\bf C})$ is positive
$\forall d \in {\bf N}$, where ${\bf I}_d$ is the $d \times d$
identity. 

CP is stronger than positivity since this latter is obtained with the
particular choice $d = 1$. The following theorem of Choi gives a
practical characterization of completely positive maps~\cite{cho},
saying that checking $d = n$ is enough:

{\bf Theorem 2.} The map $\gt: \mn \rightarrow \mn$ is completely
positive if and only
if the map $\git: \mmn \rightarrow \mmn$ is positive.

The standard form of completely positive maps was obtained by Stinespring:

{\bf Theorem 3.} $\gt$ is CP if and only if there is a set of bounded
operators $\{ V_i^{(t)} \}$ such that

\begin{equation}
\label{stine}
\gt [\rho_S] = \sum_i V_i^{(t)\dagger} \rho_S V_i^{(t)} \quad
\forall \rho_S \in \mn.
\end{equation}

We claim that positivity is not enough for a correct interpretation of
the quantum formalism. Indeed, consider the time-evolution of the 
composite system S + ${\rm S}_n$, where S is the system of interest 
evolving under $\gt$
and ${\rm S}_n$ is an arbitrary $n-$level system (where $n =$ dim
${\cal H}_S$). Suppose in particular that ${\rm S}_n$ is characterized
by the null Hamiltonian $H_n = 0$ and that S and ${\rm S}_n$ do not
interact, $H_I = 0$, then the time-evolution of S + ${\rm S}_n$ is
given by $\git$. In order to preserve the positivity of the
eigenvalues of a statistical operator representing S + ${\rm S}_n$,
this dynamics must be positive for any initial state. But, according
to Theorem 2, this is a necessary and sufficient condition for
$\gt$ being completely positive.

In different terms, if $\gt$ is only positive and not completely
positive, then there surely exists an entangled 
state $\rho$ of S + S$_n$ such
that at some $t$ $\gamma_t \otimes {\bf I}_n[\rho]$ has negative
eigenvalues and cannot be interpreted as a state.

{\bf Definition 5.} Given a composite system ${\rm S}_1 + {\rm S}_2$,
the state $\rho$ representing it is called {\it separable} if and only
if $\rho = \sum_{i} p_{i} \rho_i^1 \otimes \rho_i^2$, where
$\rho_i^1$, $\rho_i^2$ are statistical operators characterizing
systems ${\rm S}_1$, respectively ${\rm S}_2$, and $p_i \geqslant 0$,
$\sum_i p_i = 1$.

The only states in S + ${\rm S}_n$ that could exhibit an unphysical
evolution (i.e. negative eigenvalues) are entangled states, since
$\git$ certainly preserves the positivity of separable states
when $\gamma_t$ is positivity-preserving. 
Then, whereas positivity guarantees the physical consistency
of evolving states of single systems, CP prevents inconsistencies in
entangled composite systems.

It is possible to give a general characterization of completely
positive maps having the semigroup structure~\cite{gor}; 
instead, positivity has so far not got a complete characterization. 

{\bf Theorem 4.} The time-evolution generated by~(\ref{markov}) with
$L_D$ as in~(\ref{kossak}) consists of completely positive maps if and
only if the matrix $C$ defined in~(\ref{kossak}) is positive definite.

Note that the time-evolution of a closed system, eq.~(\ref{closed}),
is completely positive since it has the Stinespring form.


\section{Factorized dynamics}
\label{s3}

In this section we consider factorized Markovian dynamics for 
bipartite systems S = ${\rm S}_1 + {\rm S}_2$. 
Both subsystems are assumed to be described by
$n-$dimensional Hilbert spaces ${\cal H}_{S_1}$ and ${\cal H}_{S_2}$;
their time-evolutions are assumed to be semigroups 
of linear maps denoted by $\{ \gt^{(1),(2)},t \geqslant 0 \}$ and the
dynamics of the whole system to be given by $\{ \ggtot,t \geqslant 0
\}$. We have seen that $\gto \otimes {\bf I}_n$ positive implies
$\gto$ completely positive and the same holds for ${\bf I}_n \otimes
\gtt$. We now study what conditions the request that $\ggtot$ be 
positive puts on $\gamma_t^{(1),(2)}$. This question is justified since
there are physical situations, subject to experimental investigation,
characterized by dissipative Markovian evolutions in the factorized
form~\cite{raf}. 

Quantum open systems can be thought of as being in contact with a heat
bath at some temperature $T$; if they are bipartite, not interacting
among themselves and only weakly with the environment, their joint
time-evolution may be modeled by $\ggt$. However, due to local
fluctuactions, it may be that their dynamics be as if they were in
contact with baths at slightly different temperatures so that a
realistic description of the evolution is given by $\gt \otimes 
\gamma_t^{\varepsilon}$ with $\gamma_t^{\varepsilon}$ a perturbation
of $\gt$.

It is thus possible to give more physical flavour to the rather
abstract notion of CP, offering stronger evidences to its necessity in
the description of open systems dynamics.

The generator of $\ggtot$ is given by $L = L_1 \otimes {\bf I}_n +
{\bf I}_n \otimes L_2$, where $L_{1,2}$ generate the elementary
evolutions $\gamma_t^{(1),(2)}$. Without loss of generality we neglect any
Hamiltonian contribution in these generators (such contribution is in
fact completely irrelevant for the following discussion), then they
are expressed as in~(\ref{kossak}) with self-adjoint 
$n \times n$ matrices $C_1 =
[c_{ij}^{(1)}]$ and $C_2 = [c_{ij}^{(2)}]$. The generator $L$ is thus 

\begin{equation}
\begin{split}
\label{genfac}
L[\rho_S] = \sum_{i,j = 1}^{n^2 - 1} \Bigl(
&c_{ij}^{(1)} \Bigl[F_i \otimes {\bf I}_n \rho_S F_j^{\dagger} \otimes {\bf
I}_n - \frac{1}{2} \Bigl\{ F_j^{\dagger} F_i \otimes {\bf I}_n, \rho_S 
\Bigr\}\Bigr] + 
\\ 
+ &c_{ij}^{(2)} \Bigl[{\bf I}_n \otimes F_i \rho_S {\bf I}_n \otimes
F_j^{\dagger} - \frac{1}{2} \Bigl\{ {\bf I}_n \otimes F_j^{\dagger} F_i,
\rho_S \Bigr\}\Bigr] \Bigr)
\end{split}
\end{equation}

\noindent
and the corresponding $2n \times 2n$ matrix $C$ can be represented in
block diagonal form 

\begin{equation}
\label{genfac2}
C = \left( 
\begin{array}{cc}
C_1 & 0 \\
0 & C_2
\end{array}
\right).
\end{equation}

If both the subsystems do evolve under the same time-evolution $\gt$,
a strong equivalence can be obtained~\cite{ben3}.

{\bf Theorem 5.} If $\{ \gt, t \geqslant 0 \}$ is a 
semigroup of linear maps over the states of $\mn$, then the semigroup
$\{ \ggt, t \geqslant 0 \}$ of linear maps over the states of $\mmn$
is positivity-preserving if and only if $\{ \gt, t \geqslant 0 \}$ is
made of completely positive maps.

This result offers a new approach to CP, without reference to
arbitrary $n-$level systems usually introduced in order to give a
physical interpretation of CP (see the discussion in the previous
section). It turns out that if $\gt$ is positive but not completely
positive, there exists an entangled state in $\mmn$ mapped by $\ggt$ 
into a non-positive operator.

Does this result hold when the two time-evolutions $\gto$ and $\gtt$
are different? The answer is no, as we will shortly prove by a
counterexample, i.e. a positive factorized map $\ggtot$ with $\gtt$
violating CP. We set $n = 2$ (that is, dim${\cal H}_{S_1}$ = dim${\cal
H}_{S_2}$ = 2) and define 

\begin{equation}
\label{kossak12}
C_1 = \left(
\begin{array}{ccc}
1 & 0 & 0 \\
0 & 1 & 0 \\
0 & 0 & 1
\end{array}
\right), \quad
C_2 = \left(
\begin{array}{ccc}
1 & 0 & 0 \\
0 & -1 & 0 \\
0 & 0 & 1
\end{array}
\right).
\end{equation}

Using the coherence-vector representation for $\rho_S \in \mt$,

\begin{equation}
\label{cohvec}
\rho_S = \sum_{\mu = 0}^3 \rho^{\mu} \sigma_{\mu}, \quad
\rho^{\mu} \in {\bf R}, \quad \mu = 0, ... 3
\end{equation}

\noindent
where $\{ \sigma_i, i = 1, 2, 3 \}$ are the Pauli matrices and
$\sigma_0 = {\bf I}_2$, we get the action of the elementary
time-evolutions on $\rho_S \in \mt$:

\begin{equation}
\label{elementary}
\begin{split}
\gto [\rho_S] &= \rho^0 \sigma_0 + e^{-4 t}(\rho^1 \sigma_1 + \rho^2
\sigma_2 + \rho^3 \sigma_3), \\
\gtt [\rho_S] &= \rho^0 \sigma_0 + \rho^1 \sigma_1 + e^{-4 t} \rho^2
\sigma_2 + \rho^3 \sigma_3.
\end{split}
\end{equation}

Since in $2$ dimensions $\rho_S$ is positive if and only if det$(\rho_S) \geqslant
0$, we see that $\gto$ and $\gtt$ are positivity-preserving. Indeed
det$(\rho_S) \geqslant 0$ if and only if $\sum_{i = 1}^3 [\rho^i (t)]^2
\leqslant 1/4$, but $\sum_{i = 1}^3 [\rho^i (0)]^2 \leqslant 1/4$ (since
we start with a positive operator) and from~(\ref{elementary}) any
component
$\rho^i$ is fixed or exponentially decreasing in time under both
$\gto$ and $\gtt$. Then det$(\gt^{(1),(2)} 
[\rho_S (t)]) \geqslant 0 \,\,\forall
t \geqslant 0$ and finally $\gto$, $\gtt$ are positive.

From Theorem 4 we deduce that $\gto$ is completely positive, $\gtt$ is
not ($C_1 \geqslant 0, C_2 \ngeqslant 0$). 

Consider now $\ggtot$. 
Since $C_2$ as in~(\ref{kossak12}) is not positive, the evolution is not
completely positive\footnote{Note that an arbitrary $\ggtot$ is 
completely positive if and only if $\gto$ and $\gtt$ are completely 
positive.}. 
Is it positive? In order to prove that it is indeed so, we can
restrict our attention to its action on pure states in $\mmt$. In
fact, if a linear time-evolution is positive on pure states, it must
be positive also on mixed states. Then $\rho_S = \vert \psi \rangle
\langle \psi \vert$, with $\vert \psi \rangle \in {\cal H}_{S_1}
\otimes {\cal H}_{S_2}$. It is convenient to use the Schmidt
decomposition

\begin{equation}
\label{schmidt}
\vert \psi \rangle = \sum_{i = 1, 2} \sqrt{\mu_i}\, \vert \phi^{(1)}_i
\rangle \otimes \vert \phi^{(2)}_i \rangle, \quad
\vert \phi^{(1)}_i \rangle \in {\cal H}_{S_1}, 
\vert \phi^{(2)}_i \rangle \in {\cal H}_{S_2}, \quad
\langle \phi_i^{(a)} \vert \phi^{(a)}_j \rangle = \delta_{ij}, \; a = 1, 2
\end{equation}

\noindent
with $\mu_{1,2} \geqslant 0$, $\mu_1 + \mu_2 = 1$. Then,
since~(\ref{elementary}) holds in general for any matrix in $\mt$ (and
not only for density matrices), we
can compute $\rho_S (t) = \ggtot [\rho_S]$; writing $\mu = \mu_1$ and
defining $\alpha$, $\varphi$ such that $\langle \phi^{(2)}_1 \vert
\sigma_2 \vert \phi^{(2)}_1 \rangle = \cos{\alpha}$, $\langle \phi^{(2)}_2
\vert \sigma_2 \vert \phi^{(2)}_1 \rangle = e^{i \varphi} \sin{\alpha}$,
we get

\begin{equation}
\label{evoluz}
\rho_S (t) = e^{-4 t} \rho_S + \frac{1}{2} (1 - e^{-4 t}) 
\left[ \sigma_0 \otimes \left(
\begin{array}{cc}
\mu & 0 \\
0 & 1 - \mu
\end{array}
\right) -
Z_t \otimes \sigma_2 \right]
\end{equation}

\noindent
where

\begin{equation}
\label{zeta}
Z_t = \left(
\begin{array}{cc}
\frac{1}{2} \cos{\alpha} \, (2 \mu - 1 + e^{-4 t}) &
e^{i \varphi} \sin{\alpha} \, e^{-4 t} \sqrt{\mu (1 - \mu)} \\ \\
e^{-i \varphi} \sin{\alpha} \, e^{-4 t} \sqrt{\mu (1 - \mu)} &
\frac{1}{2} \cos{\alpha} \, (2 \mu - 1 - e^{-4 t})
\end{array}
\right).
\end{equation}

After diagonalization of $Z_t$, we can compute the doubly degenerate
eigenvalues of the second term in the r.h.s. of~(\ref{evoluz}):

\begin{equation}
\label{eigen}
z_{\pm} (t) = \frac{1}{4} (1 - e^{-4 t}) \left(1 \pm \sqrt{1- (1 -
e^{-8 t})[1 - \sin^2{\alpha}(1 - 2 \mu)^2]}\right);
\end{equation}

\noindent
they are both positive for any time $t$ as a consequence of $\mu
\in [0, 1]$, $\sin{\alpha}\in [-1, 1]$. $\rho_S (t)$ is the sum of two
positive definite matrices, then it is positive for any time $t$ and for
any initial state $\rho_S$ and $\ggtot$ is positivity-preserving
$\forall t \geqslant 0$.

Thus we have exhibited an explicit example of a positive map $\ggtot$ 
whose constituent maps $\gto$, $\gtt$ are not both completely positive.

Nevertheless, we will shortly show that, if $\gtt$ slightly differs 
from $\gto$, then $\gamma_t^{(1),(2)}$ both completely positive is a
necessary condition
for the positivity of $\ggtot$. A preliminary result is the following.

{\bf Lemma 1.} Let the semigroup $\{ \ggtot , t \geqslant 0 \}$ 
over the states of $\mmn$ have $\gto$, $\gtt$ generated by 
$L_1$, $L_2$ respectively; it consists
of positivity-preserving maps only if $C_1 + C_2 \geqslant
0$, where $C_1$, $C_2$ are the matrices of coefficients characterizing
the generators $L_1$, $L_2$ as in~(\ref{kossak}).

{\it Proof:} From positivity preservation it follows that

\begin{equation}
\label{lemma1}
{\cal G}_{\phi, \psi}(t) = \langle \phi \vert (\ggtot) [\vert \psi
\rangle \langle \psi \vert] \vert \phi \rangle \geqslant 0 \quad
\forall \vert \phi \rangle , \vert \psi \rangle \in {\bf C}^n \times
{\bf C}^n.
\end{equation}

Choosing $\langle \phi \vert \psi \rangle = 0$ it must be ${\cal
L}_{\phi, \psi} = d{\cal
G}_{\phi, \psi}(t)/dt\vert_{t = 0} \geqslant 0$ and thus 

\begin{equation}
\label{lemma2}
{\cal L}_{\phi, \psi} = \langle \phi \vert (L_1 \otimes {\bf I}_n +
{\bf I}_n \otimes L_2)[\vert \psi \rangle \langle \psi \vert] \vert
\phi \rangle \geqslant 0.
\end{equation}

Let $\{ \vert j \rangle , j= 1, ... , n\}$ an orthonormal basis of
${\bf C}^n$, then

\begin{equation}
\label{lemma3}
{\cal L}_{\phi, \psi} = \sum_{i,j = 1}^{n^2 - 1} \Bigl\{ c_{ij}^{(1)} \Bigl[\Tr
(\Psi \Phi^{\dagger} F_i) \Tr (\Phi \Psi^{\dagger} F_j^{\dagger})\Bigr] +
c_{ij}^{(2)} \Bigl[\Tr ((\Phi^{\dagger} \Psi)^T F_i) \Tr ((\Psi^{\dagger}
\Phi)^T F_j^{\dagger})\Bigr] \Bigr\} \geqslant 0
\end{equation}

\noindent
where $C_1 = [c_{ij}^{(1)}]$, $C_2 = [c_{ij}^{(2)}]$, $\{ F_i, i = 1, ... ,
n^2 - 1\}$ are the traceless matrices appearing in Theorem 1 and $\Psi
= [\psi_{ij}]$, $\Phi = [\phi_{ij}]$ are
the matrices of the
coefficients of the expansion of $\vert \psi \rangle $ and $\vert \phi
\rangle $ on the basis $\{ \vert j \rangle \otimes \vert
k \rangle ; j,k = 1, ... , n \}$. Because of the orthogonality between
$\vert \psi \rangle $ and $ \vert \phi \rangle $, $\Tr (\Phi
\Psi^{\dagger}) = 0$.

Now, for any $\vert \xi \rangle = (\xi_1, ... \xi_{n^2 - 1})^T
\in {\bf C}^{n^2 - 1}$, consider the traceless $n \times n$ matrix $W
= \sum_{i = 1}^{n^2 - 1} \xi_i F_i$ and impose $W = \Phi
\Psi^{\dagger}$. Since $W$ and its transpose $W^T$ are similar to each
other~\cite{gel}, 
define $\Phi$ such that $\Phi^{-1} W \Phi = W^T$. $\Psi$ is
then fixed by $\Psi^{\dagger} \Phi = W^T$ and equation~(\ref{lemma3})
becomes 

\begin{equation}
\label{lemma4}
{\cal L}_{\phi, \psi} = \sum_{i,j = 1}^{n^2 - 1} (c_{ij}^1 + c_{ij}^2)
\xi_i^* \xi_j \geqslant 0
\end{equation}

\noindent
that is $\langle \xi \vert (C_1 + C_2) \vert \xi \rangle \geqslant 0$
$\forall \vert \xi \rangle \in {\bf C}^{n^2 - 1}$. Therefore $(C_1 + C_2)$
must be positive definite.

\hfill $\blacksquare$

We will now consider the situation of a bipartite system evolving
according to a semigroup of the form $\{ \ggtote, t \geqslant 0 \}$
where $\gamma_t^{\varepsilon}$ is a perturbation of $\gt$, namely, if
$\gt$ is generated by $L$, $\gamma_t^{\varepsilon}$ is generated by
$L_{\varepsilon} = L + \varepsilon \, \Lambda$ where $\varepsilon \in
{\bf R}$ and $\Lambda$ acts on $\mn$. 

{\bf Theorem 6.} 
In the setting sketched above, there always exists
an interval $I_0 = [0, \varepsilon_0]$, such that
$\forall \varepsilon \in I_0$, the set $\{ \ggtote, t \geqslant 0 \}$ 
is made of positivity-preserving maps if and only if the semigroups
$\{ \gt, t\geqslant 0 \}$ and $\{ \gamma_t^{\varepsilon}, t \geqslant
0 \}$ are completely positive.


{\it Proof:} The if part is trivial, we will prove only the necessity.

To start with, we show that the non-perturbed maps $\{\gt , t \geqslant
0 \}$ must be completely positive. 




 



Choosing $\varepsilon = 0$ we obtain the set $\{ \ggt, t \geqslant 0
\}$; it is made of positive maps if and only if $\{ \gt , t \geqslant
0 \}$ is made of completely positive maps (Theorem 5).

Let $C$ and $C_{\varepsilon}$ be the Hermitian 
matrices associated to $L$, $L_{\varepsilon}$ respectively 
(Theorem 1). By the previous
result and following Theorem 4, $C$ must be positive 
definite. Since $L_{\varepsilon} = L + \varepsilon \, \Lambda$, it 
follows $C_{\varepsilon} =
C + \varepsilon \, \Gamma$, where $\Gamma$ is a $(n^2 - 1)
\times (n^2 - 1)$ matrix, satisfying
$\Gamma = \Gamma^{\dagger}$.

If the semigroup $\ggtote, t \geqslant 0$ is positivity-preserving
$\forall \varepsilon \in [0, \varepsilon^{\prime}]$, by Lemma 1 
$C + C_{\varepsilon} \geqslant 0$, whence

\begin{equation}
\label{the6}
\langle \xi \vert (C + C_{\varepsilon}) \vert \xi \rangle = 
2 \langle \xi \vert C \vert \xi \rangle + \varepsilon \langle \xi
\vert \Gamma \vert \xi \rangle \geqslant 0 \quad
\forall \vert \xi \rangle \in {\bf C}^{n^2 - 1}.
\end{equation}







Then 

\begin{equation}
\label{merda}
2 \langle \xi \vert C_{\varepsilon /2} \vert \xi \rangle = 2 \langle \xi
\vert C \vert \xi \rangle + \varepsilon \langle \xi \vert \Gamma \vert
\xi \rangle \geqslant 0;
\end{equation}

\noindent
we conclude that $\langle \xi \vert C_{\varepsilon} \vert \xi
\rangle \geqslant 0$\, $\forall \varepsilon \in I_0 = [0,
\varepsilon_0]$, $\varepsilon_0 \geqslant \varepsilon^{\prime} /2$, 
$\forall \vert
\xi \rangle \in {\bf C}^{n^2 - 1}$ and then $C_{\varepsilon} \geqslant
0$. Therefore $\forall \varepsilon \in I_0$ we have not only the
positivity of the semigroup $\{\ggtote , t\geqslant 0 \}$ but also 
the complete positivity of both the maps $\{ \gamma_t^{\varepsilon}, t
\geqslant 0 \}$ and $\{ \gamma_t, t \geqslant 0 \}$.

\hfill $\blacksquare$


\section{Conclusions}
There are many physical systems whose irreversible time-evolutions are
suitably described by factorized dynamics $\ggt$. Usually, they are
bipartite systems living in the same environment and evolving under
irreversible dynamics because of their (weak) interaction with it. In
this work we have addressed the problem of a non-uniform environment,
causing the two subsystems to evolve under possibly different
time-evolutions, $\gto$ and $\gtt$.

We have shown that, while the minimal request of positivity of the
factorized dynamics $\ggtot$ does not in general imply the CP of the
constituent maps, however, they must be so if $\gtt$ is a perturbation
of $\gto$ as in the case under experimental circumstancies when
treating systems in a heat bath.

We can thus conclude that, when small temperature fluctuactions
occour, in order to avoid physical inconsistencies in the
time-evolution $\ggtot$ of bipartite systems in heat baths,
$\gto$ and $\gtt$ have to be completely positive maps.


\section*{Acknowledgments}
One of the authors (R. R.) wants to thanks the organizers for the
invitation. Work supported in part by Istituto Nazionale di Fisica
Nucleare, Sezione di Trieste, Italy.

\noindent


\end{document}